\documentstyle[12pt]{article}
\input psfig.sty

\begin{document}

\baselineskip=17pt
\pagestyle{plain}
\setcounter{page}{1}

\begin{titlepage}

\begin{flushright}
CLNS-98/1565 \\
hep-th/9806234
\end{flushright}
\vspace{10 mm}

\begin{center}
{\huge The Nonperturbative Gauge Coupling }
\vskip 3mm
{\huge  of $N=2$ Supersymmetric Theories} 
\vspace{5mm}
\end{center}
\vspace{10mm}
\begin{center}
{\large Philip C. Argyres and Alex Buchel}\\
\vspace{3mm}
{\it Newman Lab., Cornell University, Ithaca NY 14853}\\
argyres,buchel@mail.lns.cornell.edu
\end{center}
\vspace{15mm}
\begin{center}
{\large Abstract}
\end{center}
\noindent
We argue that the topology of the quantum coupling space and the
low energy effective action on the Coulomb branch of scale invariant
$N=2$ $SU(n)$ gauge theories pick out a preferred nonperturbative 
definition of the gauge coupling up to non-singular holomorphic
reparametrizations.
\vspace{1cm}
\begin{flushleft}
June 1998
\end{flushleft}
\end{titlepage}
\newpage


\newcommand{\ATMP}{Adv.\ Theor.\ Math.\ Phys.\ }
\newcommand{\JHEP}{J.H.E.P.\ }
\newcommand{\NP}{Nucl.\ Phys.\ }
\newcommand{\PL}{Phys.\ Lett.\ }
\newcommand{\PR}{Phys.\ Rev.\ }
\newcommand{\PRL}{Phys.\ Rev.\ Lett.\ }
\newcommand{\CMP}{Commun.\ Math.\ Phys.\ }
\newcommand{\be}{\begin{equation}}
\newcommand{\ee}{\end{equation}}
\newcommand{\bea}{\begin{eqnarray}}
\newcommand{\eea}{\end{eqnarray}}
\newcommand{\bz}{{\bf Z}}
\newcommand{\br}{{\bf R}}
\newcommand{\cf}{{\cal F}}

The quantum coupling space of scale invariant $N=2$ supersymmetric
gauge theories is a subset of the classical one obtained by discrete
identifications under the action of the S-duality group.  These
S-duality identifications imply an exact quantum equivalence between
classically inequivalent theories.  However, while the topology of the
quantum coupling space has an invariant meaning, its parametrization
is unambiguous only at weak coupling. Different parametrizations can
give rise to different S-duality group actions on the gauge couplings.

In one nonperturbative definition of the coupling, the S-duality group
was found to be $\Gamma^0(2)\subset SL(2,\bz)$ for the scale invariant
$SU(n)$ $N=2$ SQCD theories \cite{aps9505}.  This S-duality group is
generated by $T:\ \tau\to\tau+2$ and $S:\ \tau\to-1/\tau$ subject to
the single constraint $S^2=1$.  (We take the gauge coupling to be
$\tau={\theta\over\pi}+ i {8\pi\over g^2}$, differing by a factor of
two from the usual definition.)  A different nonperturbative
parameterization of the $SU(3)$ gauge coupling proposed in
\cite{mn9507} gives an S-duality group with the same $T$ generator,
but a different $S:\ \tau\to -4/(3\tau)$.  In the first case the
fundamental domain of the S-duality group is the region $|\tau|\ge1$
and $|{\rm Re}\tau|\le 1$ which has cusps (zero opening angle) at
$\tau=i\infty$ and $\tau=1$ and a $\bz_2$ orbifold point at $\tau=i$.
By contrast, in the second case the fundamental domain is $|\tau|\ge
2/\sqrt{3}$ and $|{\rm Re}\tau| \le 1$ with cusp at weak coupling,
$\bz_2$ orbifold at $\tau = 2i/\sqrt{3}$ and $\bz_6$ orbifold at
$\tau=1+i/\sqrt3$.  Although there exists a unique conformal map
between these two fundamental domains mapping the strong coupling cusp
of the first to the $\bz_6$ point of the second, the map is singular
there since the opening angle changes.  This is reflected in the fact
that the corresponding S-duality groups are not isomorphic as abstract
groups.

In this letter we argue that the topology of the quantum coupling
space together with the low energy effective action on the Coulomb
branch pick out a preferred nonperturbative definition of the gauge
coupling up to non-singular holomorphic reparametrizations.  We
explicitly compute the S-duality group associated with this definition
of the coupling, finding agreement with the results of \cite{m9806}
which were obtained by a different method.  We find for the finite
$N=2$ SQCD with $SU(n)$ gauge group that the S-duality group
can be presented as the subgroup of $SL(2,\br)$ generated by $T:
\tau\to\tau+2$ and
\be
S: \ \ \tau \, \to \, \left\{
\matrix{-1\over\tau, & \qquad n\ \mbox{even}\cr
-{\rm sec}^2(\pi/2n)\over\tau, & \qquad n\ \mbox{odd}\cr}\right.
\label{s}
\ee
acting on the classical coupling space $\{{\rm Im}\tau > 0\}$.

Before turning to the nonperturbative definition of the gauge
coupling, we wish to point out some properties \cite{m9806} of the
result (\ref{s}).  The S-duality group is not freely generated
by $T$ and $S$, but is subject to the constraints
\be
S^2=1
\label{firstc}
\ee
and
\be
(ST^{-1})^{2n}=1,\qquad \mbox{for}\ n\ \mbox{odd}.
\label{secondc}
\ee
For the $SU(n)$ with $n$ even the S-duality group is isomorphic to
$\Gamma^0(2)$, though this duality group is naturally enlarged to
$SL(2,\bz)$ in the $SU(2)$ case \cite{sw9408,a9706}.  For $n$ odd the
fundamental domain is defined by $|{\rm Re}\tau|\le1$ and $|\tau|
\ge\sec(\pi/2n)$ with edges identified.  This fundamental domain has
three special points: a weak coupling singularity at $\tau=i\infty$, a
$\bz_2$ orbifold point at $\tau=i\sec(\pi/2n)$, and a $\bz_{2n}$
orbifold point at $\tau=\pm 1+i \tan(\pi/2n)$.  It is an open question
whether there is an alternative characterization of the physics of at
the $\bz_{2n}$ orbifold points of the odd $n$ theories or the strong
coupling cusp points of the even $n$ theories.

We now turn to the topological definition of the coupling parameter
in the $SU(n)$ scale invariant theories.

The quantum coupling space of scale invariant $N=2$ SQCD has isolated
singularities at special couplings where the {\it whole} Coulomb
branch is singular.  Just as traversing paths around the singularities
on the Coulomb branch generate elements of the low energy
electric-magnetic (EM) duality group (reflected in monodromies of the
BPS spectrum), we argue that monodromies of the BPS spectrum around
the singularities of the quantum coupling space, $\cf$, generate the
S-duality group.

Consider the scale invariant $N=2$ supersymmetric gauge theory with
$SU(n)$ gauge group and $2n$ hypermultiplets in the fundamental
representation.  At a generic point on the Coulomb branch the gauge
group is broken to $U(1)^{n-1}$ whose effective couplings $\tau_{ij}$
form a section of an $Sp(2n-2,\bz)$ bundle on the Coulomb branch
reflecting the EM duality identifications of the low energy effective
description.  The matrix of the effective couplings was identified in
\cite{aps9505} with the the period matrix of the genus $n-1$
hyperelliptic curve $\Sigma_n$
\be 
y^2=P^2(x)-fx^{2n}
\label{su}
\ee
where $P(x)=x^n-\sum_{\ell=2}^n u_{\ell}\ x^{n-\ell}$.  The moduli
$u_{\ell}$ parametrize the Coulomb branch and $f$ is a function of the
gauge coupling $\tau$.  At weak coupling $\tau\to i\infty$, $f\sim
e^{i\pi\tau}$.  It is important to note that the coupling $f$, being a
parameter specifying the quantum field theory and not an order
parameter (vev) specifying a vacuum, cannot depend on the $u_\ell$.
Thus a ``total parameter space'' including both the coupling parameter
and the Coulomb branch vevs has the structure of a fiber bundle with
the Coulomb branch as its fibers and the coupling space $\cf$ as the
base.

The complex structure of $\Sigma_n$ degenerates whenever the
discriminant of (\ref{su}) vanishes.  At fixed coupling $f$ these
singularities of the low-energy effective action are resolved by
including in the effective description all states that become massless
there.  The charge vectors of BPS states which are massive in the
vicinity of a singularity undergo $Sp(2n-2,\bz)$ EM duality group
monodromies upon traversing closed paths in the Coulomb branch around
the singularity.

Equivalently, we can think about the same singularities as
singularities in the coupling parameter space, $\cf$, at a fixed
vacuum $u_\ell$.  We call such singularities ``specific
singularities'' since their locations depend on the specific values of
vacuum moduli.  Specific singularities are not the only singularities
in $\cf$.  The complex structure of (\ref{su}) degenerates also at the
``special singularities'' $f=f_s\equiv \{0,1,\infty\}$.  These values
of the coupling parameter are special in that the whole Coulomb branch
becomes singular whenever $f=f_s$.  Thus we can think of the quantum
coupling space $\cf$ as a three punctured sphere.  More generally, in
$N=2$ scale invariant theories with simple gauge group, the gauge
coupling $\tau$ is a section of a holomorphic line bundle over a three
punctured sphere whose structure group is identified with the
S-duality group.  One of the punctures corresponds to weak coupling
and the other two to special strongly coupled theories.  We will later
identify the monodromies around $f=0$ and $\infty$ as the $T$ and $S$
duality generators, respectively; see Fig.~1.

\begin{figure}
\centerline{\psfig{figure=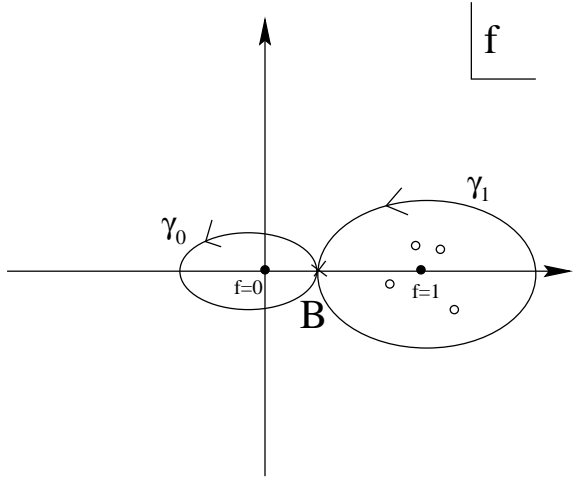,width=3truein}}
{\caption{The basis of the S-duality generators $T$ and $S T^{-1}$
is associated with the noncontractable loops $\gamma_0$ and 
$\gamma_1$. The filled dots represent ``special'' singularities 
in $\cf$, and the empty ones represent  ``specific'' singularities. 
A third special singularity is at $f=\infty$.}}
\end{figure}

As an illustration of the difference between specific and special 
singularities,  consider the scale invariant $SU(3)$ theory. 
The Coulomb branch of this theory is described by a curve $\Sigma_3$:
\be
y^2={(x^3-u_2 x -u_3)}^2-f x^6.
\label{su3}
\ee   
The complex structure of $\Sigma_3$  degenerates when the
discriminant of the right hand side of (\ref{su3}) vanishes,
\be
u_3^{10}(f-1) f^3 \left(f-{\left[1-{4 u_2^3\over 27 u_3^2}\right]}^2 
\right)=0,
\label{dsu3}
\ee
and for $f\to\infty$ with $u_\ell$ kept finite.  (In the latter case
by an appropriate rescaling of $x$ and $y$, $\Sigma_3$ reduces to the
singular curve $y^2=x^6$.)  Clearly, $f=\{0,1,\infty\}$ are always
singularities of the low-energy effective action irrespective of the
choice of the vacuum moduli $u_\ell$.  These are the special
singularities of the coupling parameter space $\cf$.  The fourth
singularity in $\cf$ is at $f=[1-(4u_2^3 / 27u_3^2)]^2$.  This
``specific'' singularity differs from the previous three in that its
position in $\cf$  depends on the choice of the Coulomb branch
vacuum.  While there are always three special singularities at fixed
positions in $\cf$ for any rank of the gauge group, the number of
specific singularities is rank dependent.  For example, the scale
invariant $SU(2)$ theory does not have specific singularities at all.
Note that the $SU(3)$ vacuum with $u_2=0$ and $u_3\ne 0$ is special in
that the specific singularity coincides with the $f_s=1$ special
singularity.

The monodromies of the BPS spectrum or low energy $U(1)^{n-1}$
couplings around singularities on the coupling space $\cf$ encode
information about the S-duality group.  But while S-duality
transformations on the gauge coupling $\tau$ must be the same for any
choice of vacuum moduli and hypermultiplet mass parameters, the
monodromies in $\cf$ around $f_s$ actually depend on this choice.
This presents a puzzle: How can the S-duality group information which
is invariant under changes in the vacuum moduli be extracted from
these monodromies?  The key to solving this puzzle is to realize that,
in principle, noncontractable loops in $\cf$ can be generators of
both the S-duality and the low energy EM duality groups.  In fact,
nontrivial loops around the specific singularities in $\cf$ have
nothing to do with the S-duality group: any such loop can be deformed
in the combined Coulomb branch and quantum coupling space to a loop
around a singularity on the Coulomb branch at a fixed value of the
gauge coupling parameter $f$.  This follows simply from the fiber
bundle structure of the combined vev plus parameter space pointed out
earlier, and the fact that by definition the specific singularities move
in the coupling space $\cf$ as we move in the Coulomb branch.  We will
call the monodromies around loops that encircle only specific
singularities in $\cf$ ``pure EM monodromies''.

So an S-duality group (and hence the coupling $\tau$ it acts on) can
be defined as the subgroup of the $Sp(2n-2,\bz)$ EM duality group
generated by the monodromies around the special singularities $f_s$ in
$\cf$.  More precisely, those monodromies will be representations of
the S-duality group generators in the low energy EM duality group
$Sp(2n-2,\bz)$ modulo the pure EM monodromies.  In this way the
S-duality group does not suffer from any ambiguity in the choice of
basis cycles around the special singularities in $\cf$ since how those
cycles are chosen to go around the specific singularities can only
change the monodromies by pure EM monodromies.  We must also mod out
the representation of the S-duality group generators in $Sp(2n-2,\bz)$
by its $GL(n-1,\bz)$ subgroup generated by simple (integral) change of
bases of the $U(1)^{n-1}$ charge lattice.  This is because the
definition of the microscopic coupling in the classical theory is
already insensitive to any choice of basis of $U(1)^{n-1}$ generators.

The S-duality group constraints (\ref{firstc}) and (\ref{secondc}) can
now be explicitly computed as follows.  We first note that there is a
choice of $SU(n)$ vacua on the Coulomb branch, namely $u_{\ell}=0$,
$\ell=2,\ldots,n-1$, and $u_n\equiv u\ne 0$, for which the step of
modding out by EM monodromies is done automatically.  This is simply
because for this choice of vacuum {\it all} the singularities in $\cal
F$ are ``special singularities''.  (This Coulomb branch submanifold
was used in \cite{ay9601,mn9601,m9806} where S-duality identifications
were analyzed from an algebraic point of view; here we just use these
vacua for calculational convenience.)  In these special vacua there is
an unbroken global $\bz_n$ discrete subgroup of the anomaly-free
$U(1)_R$ which implies that the branch points of the curve (\ref{su})
are distributed in pairs around the $n$-th roots of unity.  Choose an
overcomplete basis of cycles $\{\beta_a,\gamma^a\}$, $a=1,\ldots,n$,
as in Fig.~2; note that $\sum_a \beta_a = \sum_a \gamma^a = 0$.

\begin{figure}
\centerline{\psfig{figure=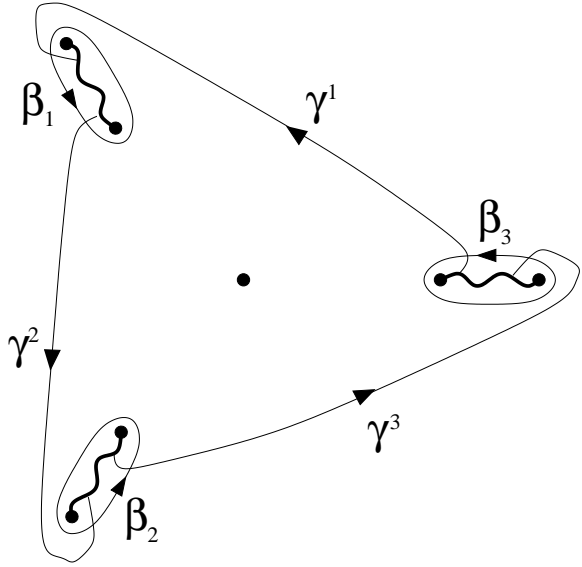,width=3truein}}
{\caption{Contours for a basis of cycles for the $SU(n)$ curve
($n=3$) on one sheet of the $x$-plane.  The thick wavy lines 
represent the cuts and the $\gamma^a$ contours close on the
second sheet.}}
\end{figure}

As $f\to e^{2\pi i} f$ for $|f|\ll 1$ (thus performing the $T$
monodromy around $f=0$) one can show by contour dragging that
\be
T:\ \ \left\{
\matrix{\beta_a & \to & \beta_a \cr
\gamma^a & \to & \gamma^a + \beta_a - \beta_{a-1} \cr}\right.
\label{Tmonod}
\ee
(where $\beta_0 \equiv \beta_n$).  Repeated $T$ monodromies
clearly do not close on the identity.  Now take $f$ close to
1.  The arrangement of cuts and cycles in Fig.~2 stays the same
except that the branch points further from the origin move off
towards $x=\infty$.  As $(f-1) \to e^{2\pi i} (f-1)$ for
$|f-1|\ll 1$ (thus performing the $ST^{-1}$ monodromy around
$f=1$) we find
\be
ST^{-1}:\ \ \left\{
\matrix{\beta_a & \to & - \beta_a +\beta_{a-1} +\gamma^a \cr
\gamma^a & \to & \beta_{a-1} \cr}\right.
\label{STmonod}
\ee
It is a simple exercise to show that for $n$ odd (\ref{secondc}) is
satisfied, while for $n$ even there is no relation.  Finally, from
(\ref{Tmonod}) and (\ref{STmonod}) we derive
\be
S^2:\ \ \left\{
\matrix{\beta_a & \to & \beta_{a-1} \cr
\gamma^a & \to & \gamma^{a-1} \cr}\right.
\label{S2monod}
\ee
This is one of the $GL(n-1,\bz)$ monodromies which implement a change
of $U(1)^{n-1}$ basis and are invisible to the microscopic coupling
$\tau$.  (\ref{S2monod}) therefore implies the constraint
(\ref{firstc}), thus completing the derivation of the S-duality group
of the $SU(n)$ scale invariant theory.

It would be interesting to extend this construction to theories with
other simple and semi-simple gauge groups.  In particular, extension
to the elliptic models of \cite{w9703} may permit a comparison with
the $SL(2,\bz)$ S-duality group of $N=4$ gauge theories.

\section*{Acknowledgments}
It is a pleasure to thank T.-M. Chiang, J. Diller, B. Greene, Z.
Kakushadze, J. Minahan, A. Shapere, G. Shiu, H. Tye, E. Witten, and
P. Yi for helpful comments and discussions.  PCA thanks the University
of Cincinnati Physics Department for its hospitality.  This work is
supported in part by NSF grant PHY-9513717.  The work of PCA is
supported in part by an A.P. Sloan fellowship.

\end{document}